\documentclass{article}
\usepackage{iclr2027_conference,times}
\usepackage[T1]{fontenc}
\usepackage[utf8]{inputenc}
\usepackage{amsmath,amssymb}
\usepackage{graphicx,booktabs,tabularx,array,longtable}
\renewcommand{\arraystretch}{0.96}
\usepackage{float}
\ifdefined\pdfminorversion
\else
\AtBeginDvi{}
\fi
\usepackage{microtype}
\usepackage{hyperref}
\usepackage{xurl}
\hypersetup{colorlinks=true,citecolor=blue,urlcolor=blue,linkcolor=blue,
  pdftitle={AutoSynth: Learning to Generate Editable Synthesizer Programs from Audio and Text},
  pdfauthor={Tristan Wu, Daniel Chin, Liwei Lin, Junan Zhang, Gus Xia}}

\title{AutoSynth: Learning to Generate Editable Synthesizer Programs from Audio and Text}
\author{Tristan Wu$^1$ \quad Daniel Chin$^2$ \quad Liwei Lin$^2$ \quad Junan Zhang$^3$ \quad Gus Xia$^4$\\[4pt]
{\normalfont\small $^1$ The Hong Kong University of Science and Technology (Guangzhou)}\\
{\normalfont\small $^2$ New York University Shanghai}\\
{\normalfont\small $^3$ The Chinese University of Hong Kong, Shenzhen}\\
{\normalfont\small $^4$ Mohamed bin Zayed University of Artificial Intelligence}}
\iclrfinalcopy
\begin{document}
\maketitle
\lhead{arXiv preprint}
\begin{abstract}
Audio generation models can translate natural-language descriptions into sound, but their outputs are typically waveforms. Their audio quality is constrained by audio compression, and their outputs do not readily support direct edits to notes, timbral parameters, or modulation relationships. We present AutoSynth, which represents MIDI performance events, fixed synthesizer parameters, and variable-length modulation routes as a unified sequence for a synthesizer, and learns their dependencies with an audio-conditioned autoregressive model. A single model supports both tasks. Given reference audio, the model directly predicts a synthesizer program; given text, it uses a pretrained audio generation model and converts the generated audio into a program. Training consists of two stages: supervised learning on large-scale audio--program pairs automatically constructed from a small set of native presets, followed by group-relative policy optimization with a mixed reward combining semantic similarity, pitch-related features, acoustic similarity, and sound usefulness. The pipeline requires neither paired text--target-program annotations nor a differentiable synthesizer. Experiments show that AutoSynth produces complete, editable synthesizer programs and achieves competitive results in both synthesizer inversion and text-driven generation.
Audio demos and source code are available at \url{https://auto-synth.github.io/}.
\end{abstract}

\section{Introduction}

End-to-end audio generation has long faced two challenges. First, generated audio quality remains affected by representation compression and waveform reconstruction. Researchers have moved from mel spectrograms toward direct waveform compression and continued to improve audio codecs, but a trade-off remains between high compression ratios and fine-detail fidelity \citep{zeghidour2021soundstream,defossez2022encodec,kumar2023dac}. Second, generated outputs offer limited editability: modifying an individual note, filter, or modulation relationship is difficult compared with conventional music production. Generating editable content with AI inside digital audio workstations (DAWs) has therefore attracted increasing attention \citep{elkins2025dawzy,lu2026webdaw}.

Music and sound-effect generation continues to advance \citep{copet2023musicgen,evans2026stable,hadjeres2026woosh}, yet generating content that users can directly edit in DAWs remains challenging. Computer vision and graphics have extensively explored 3D asset generation from text or reference images \citep{xiang2025trellis,meshy} and language-guided scene editing \citep{huang2024blenderalchemy}. In audio, jointly generating performance events, timbral parameters, and modulation structures remains comparatively underexplored. This requires models to determine how a sound is performed, how its timbre is configured, and how these controls evolve over time.

We argue that one important reason for this gap is that existing symbolic representations usually capture only part of music production. MIDI and ABC can express pitch and timing \citep{lu2023musecoco,yuan2024chatmusician}, but these alone do not determine the resulting sound. Much of the information involved in music production resides in synthesizer parameters and their modulation relationships. For sounds within a synthesizer's expressive range, MIDI, parameters, and modulation structure jointly form a compact, executable description of sound.

Synthesizer parameters are also an interpretable and editable symbolic language that people have used for many years: a representation that has evolved through sound-making practice. Their synthesis rules are consequently intertwined with prior knowledge about the world and sound-design aesthetics. For example, a sine wave contains no additional harmonics and therefore sounds pure, whereas flanging can produce a metallic timbre. Learning these controls involves not only fitting waveforms, but also learning the aesthetics of sound creation. Enabling AI to learn synthesizer parameters is therefore an important step toward improving its musical understanding and sound-design capabilities.

However, parameter spaces are not naturally amenable to model training. Synthesizers expose different control interfaces; parameters depend on MIDI and on one another; and customizable modulation routes (e.g., a low-frequency oscillator controlling pitch to create periodic variation) make program structure variable. Moreover, different programs can produce similar sounds \citep{hayes2025permutations}, while native plugins generally do not support direct backpropagation. Limited paired data further increases the difficulty of training.

Prior work has explored differentiable synthesis \citep{engel2020ddsp,yang2023whitebox,uzrad2024diffmoog}, black-box synthesizer inversion \citep{hayes2025permutations,shin2025synthrl,wu2026ddsynth}, and text-driven parameter generation \citep{cherep2024ctag}. Nevertheless, jointly modeling performance events, timbral parameters, and variable-length modulation structure, while making these methods applicable to practical settings, remains an open direction.

We present AutoSynth, which targets Helm and uses an autoregressive decoder to generate MIDI, synthesizer parameter values, and modulation routes in sequence, forming a complete, executable synthesizer program. An audio encoder trained jointly with the decoder extracts conditioning information. For text input, Stable Audio 3 first generates reference audio, which is then inverted by the same model. Starting from a few hundred native presets, we automatically construct supervised data and apply rendering-based group-relative policy optimization (GRPO) post-training \citep{shao2024deepseekmath}. This transfers audio generation capabilities into editable synthesizer controls without requiring paired text--target-program annotations.

Our contributions are:
\begin{enumerate}
\item \textbf{An autoregressive representation jointly modeling MIDI, synthesizer parameters, and modulation structure.} Performance events, fixed parameters, and variable-length routes are organized into a unified sequence, enabling dependency learning in a complex joint control space and providing a modeling basis for extension to other synthesizers.
\item \textbf{A two-stage learning pipeline without paired text--target-program annotations.} Large-scale supervised data are constructed automatically from a limited preset collection, and GRPO with mixed audio rewards uses sound similarity and usefulness feedback to optimize program generation.
\item \textbf{An audio- and text-driven framework sharing one inversion model.} The framework supports both synthesizer inversion and text-driven generation through an intermediate generated waveform, producing outputs that users can edit directly.
\end{enumerate}

\begin{figure}[t]
\centering
\includegraphics[width=\linewidth,trim=0 14bp 0 10bp,clip]{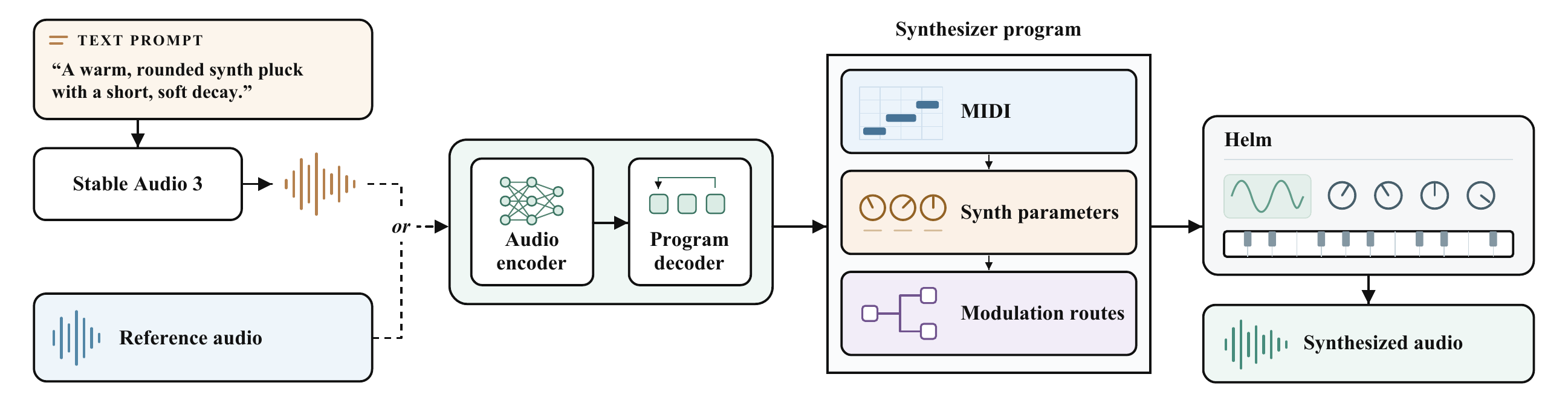}
\caption{Overview of AutoSynth.}
\label{fig:overview}
\end{figure}

\section{Related Work}

\paragraph{Audio generation and editing.}
Audio generation has advanced along two main directions: autoregressive modeling of discrete audio tokens \citep{borsos2022audiolm,copet2023musicgen} and denoising-based generation in continuous audio representations \citep{liu2023audioldm,evans2026stable,hadjeres2026woosh}. Audio quality and output editability remain important challenges. To improve quality, studies have refined audio compression and reconstruction through multi-stage residual quantization that preserves more detail at higher bitrates \citep{zeghidour2021soundstream,defossez2022encodec}, or improved quantization and decoder architectures, reconstruction losses, and adversarial training \citep{kumar2023dac}. Continuous audio compression also benefits from phase-aware reconstruction losses and improved discriminators \citep{parker2026same}. To improve editability, another line of work supports post-generation changes through text-guided sound transformations and inpainting \citep{liu2023audioldm,lelan2024melodyflow,evans2026stable}. However, compression rate and fine-detail fidelity still trade off, and audio-level edits do not map directly to the notes, timbral parameters, and modulation relationships used in music production. Translating generation capabilities into directly manipulable creative content remains an open direction.

\paragraph{Generation and procedural creation in graphics.}
Graphics research addresses not only visual outputs but also underlying assets that remain editable throughout the creative process. Prior work constructs 3D models from text or reference images \citep{poole2022dreamfusion,xiang2025trellis,meshy}, uses vision-language models to assist scene editing \citep{huang2024blenderalchemy,ricouard2026astra}, and automates character rigging \citep{xu2020rignet} to support further creation. These directions share the idea of controlling rendered outputs by generating and editing underlying structures, paralleling sound control through synthesizer programs and motivating our exploration.

\paragraph{Text-driven generation.}
Compared with directly generating waveforms, symbolic representations offer more explicit points of control. Prior studies convert text into symbolic music that can be edited at the note level, although notes alone do not determine timbre or synthesis \citep{lu2023musecoco,yuan2024chatmusician}. To control sound characteristics, other work uses synthesizer-parameter search guided by text--audio similarity, enabling sound design without paired text--parameter annotations \citep{cherep2024ctag}. However, jointly generating performance events, timbral parameters, and variable-length modulation structures from natural language, while modeling their dependencies to produce programs suitable for practical creation, remains an open direction.

\paragraph{Synthesizer inversion.}
Synthesizer inversion seeks control parameters that reproduce a reference sound, with methods progressing from black-box search to neural prediction \citep{horner1993genetic,heise2009cloning,levaillant2021vae,wu2026ddsynth}. Learning audio--parameter mappings and timbral latent spaces enables parameter estimation and timbre control \citep{esling2019flow,chen2022sound2synth,levaillant2024spinvae}. Since different parameter configurations can produce similar sounds, subsequent work uses parameter permutation symmetries to model multiple valid solutions while jointly predicting MIDI information \citep{hayes2025permutations}. To move beyond parameter error as a measure of inversion quality, one direction constructs differentiable synthesis processes and directly optimizes audio losses \citep{engel2020ddsp,uzrad2024diffmoog}; another uses reinforcement learning with native-plugin rendering feedback, without backpropagating through the synthesizer \citep{shin2025synthrl,wu2026ddsynth}. When the control space also includes performance events and variable-length modulation routes, jointly learning these structures and their dependencies, while optimizing the actual rendered sound, remains a research challenge.

\paragraph{Software synthesizers and control spaces.}
The choice of synthesizer determines both the range of expressible sounds and the control structures a model must learn. Prior work has considered the classic DX7-style FM synthesizer Dexed and the more feature-rich modern synthesizer Surge XT \citep{dexed,surge}, whose parameter definitions and synthesis rules differ. We choose Helm because its controls resemble those of commonly used software synthesizers, while its open-source implementation and 274 factory presets facilitate data construction and native rendering \citep{helm}. It is thus suitable for studying the joint modeling of performance, timbre, and modulation structure.

\section{Methodology}
\subsection{A Unified Audio- and Text-Driven Framework}

AutoSynth consists of an audio encoder and an autoregressive program decoder, modeling the conditional distribution $p_\theta(y\mid x)$ from reference audio $x$ to synthesizer program $y$. The encoder converts a waveform into a sequence of conditioning features, from which the decoder predicts a program. A native synthesizer then executes the program to render sound.

Synthesizer inversion directly uses the input audio as $x$. For text-driven generation, a frozen Stable Audio 3 Medium model \citep{evans2026stable} first converts the description into a waveform, which is then used as $x$. Both paths share the same encoder, decoder, and renderer; text is not directly fed into the program decoder.

During SFT and post-training, the audio encoder and program decoder are jointly updated, and their trainable parameters are collectively denoted by $\theta$. Stable Audio 3 remains frozen. The native synthesizer is used only for rendering and does not participate in backpropagation.

\subsection{Structured Autoregressive Synthesizer Modeling}

We serialize MIDI performance events $M$, fixed parameters $P$, and variable-length modulation routes $R$ into a program $y$, in that order. MIDI comprises pitch, velocity, onset, duration, and optional controller events; each route consists of a modulation source, destination, and amount. The model predicts the program autoregressively conditioned on audio $x$, with the following component-wise factorization:
\begin{equation}
p_\theta(M,P,R\mid x)=p_\theta(M\mid x)\,p_\theta(P\mid M,x)\,p_\theta(R\mid M,P,x).
\label{eq:ar}
\end{equation}
\begin{figure}[!b]
\centering
\includegraphics[width=\linewidth]{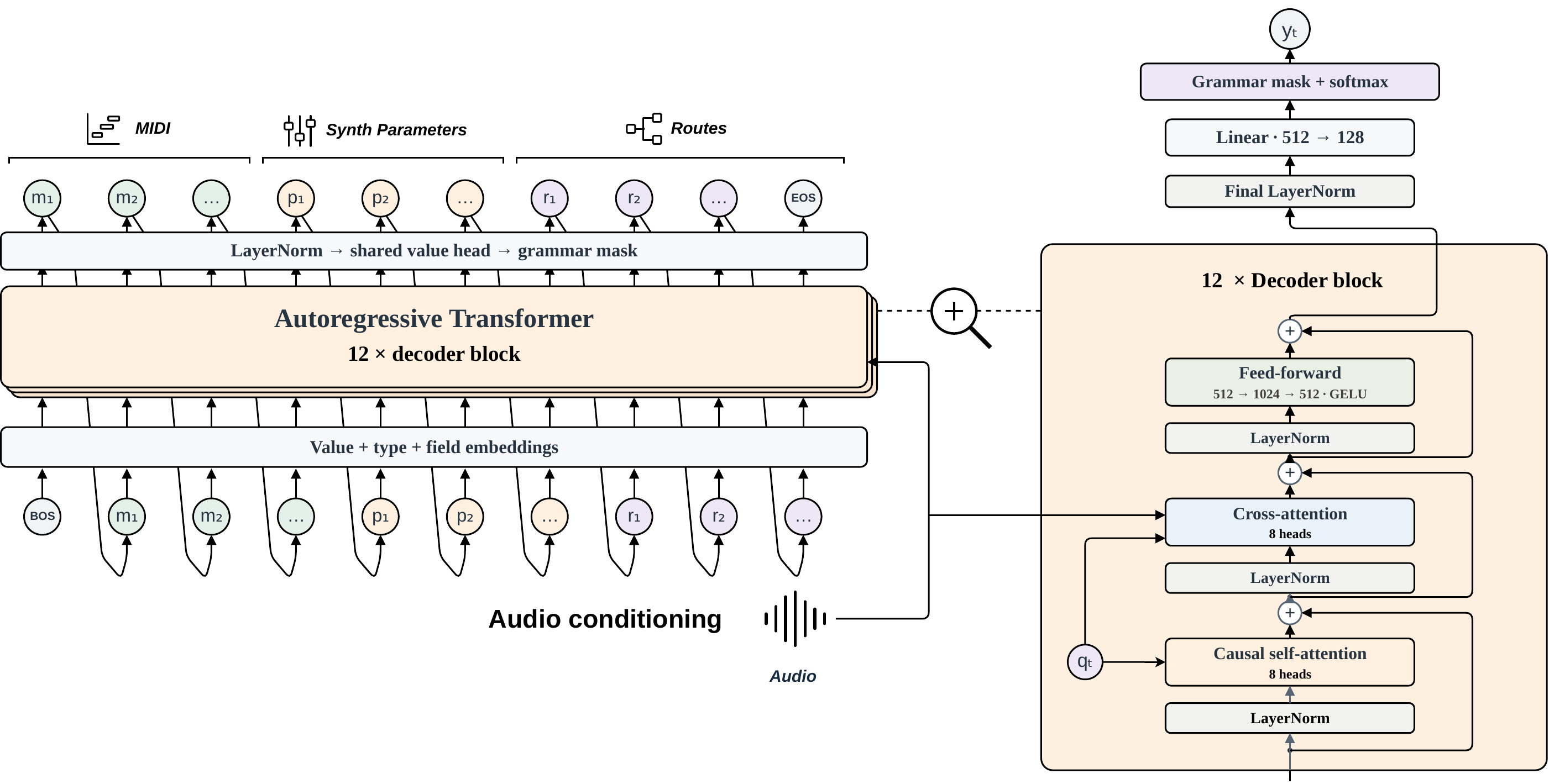}
\caption{Autoregressive program generation and decoder architecture.}
\label{fig:decoder}
\end{figure}

Generating MIDI, fixed parameters, and routes sequentially allows timbral configuration to depend on performance information, and modulation structure to depend on both previously generated components. Continuous values are discretized into 128 bins, while native discrete parameters retain their original number of categories. Routes follow a canonical ordering, with an end marker determining their number, preventing different permutations of the same route set from becoming distinct supervision targets. The complete sequence format, event limits, and fixed-parameter order are given in Appendix~\ref{app:ordering}.

The audio frontend computes log-mel spectrograms separately for the left and right channels, then produces a conditioning sequence using convolutional feature extraction and a 12-layer Transformer encoder. The program decoder contains 12 Transformer layers, each using causal self-attention over preceding tokens and cross-attention to the audio conditioning \citep{vaswani2017attention}. Model size and input-processing settings appear in Appendix~\ref{app:hyperparameters}.

Token representations contain value, type, and field information. The three program components are parts of a single sequence generated by the same decoder, and all positions share one value-classification output layer. During generation, the current grammatical position constrains the set of legal values.

\subsection{Supervised Learning from Synthesized Pairs}

Starting from native presets, we apply constrained sampling and augmentation to synthesizer parameters, modulation routes, and MIDI performances, and render audio--program pairs $(x,y)$ with the native plugin \citep{braun2021dawdreamer}. Presets provide sound-design priors, while parameter and structural augmentation expand the learnable control space.

SFT uses teacher forcing to predict the next token from the ground-truth prefix. To prevent frequently occurring fields from dominating training, we first average cross-entropy within each semantic field, then average over the fields present in the batch:
\begin{equation}
\mathcal L_{\mathrm{SFT}}=
\mathbb E_{\mathcal B}\left[
\frac{1}{|\mathcal F_{\mathcal B}|}
\sum_{f\in\mathcal F_{\mathcal B}}\mathcal L_{\mathrm{CE}}^{(f)}
\right].
\label{eq:sft}
\end{equation}
Here, $\mathcal B$ denotes a minibatch sampled from the supervised training data, and $\mathbb E_{\mathcal B}$ averages over such minibatches. $\mathcal F_{\mathcal B}$ is the set of type--field combinations present in the batch, and $\mathcal L_{\mathrm{CE}}^{(f)}$ is the mean cross-entropy for field $f$. Ordered numerical fields use Gaussian-smoothed labels; other fields use hard labels. Both the audio encoder and program decoder are updated in this stage. Label-smoothing and optimization hyperparameters appear in Appendix~\ref{app:hyperparameters}.

\subsection{Mixed-Reward Optimization from Rendering Feedback}

Parameter supervision encourages reproduction of a given program, but does not directly measure whether its sound matches the target \citep{han2024perceptual}. Because multiple programs can produce similar sounds, we further optimize the model using native plugin renders. This stage does not require synthesizer parameter annotations for the target audio.

\paragraph{Mixed targets and rewards.}
Post-training randomly mixes Stable Audio 3-generated audio and in-domain Helm training audio at a ratio of $5:3$. For each target $x$, the model generates $G$ candidate programs $y_i$, renders them as $\hat x_i=\mathcal S(y_i)$ using the native synthesizer $\mathcal S$, and computes a weighted sum of eight reward components:
\begin{equation}
r_i=\sum_{k=1}^{8}\lambda_k s_k(x,c,\hat x_i).
\label{eq:reward}
\end{equation}
The components are CLAP distance, Basic Pitch hidden-feature distance, AudioBox CE and CU, wMFCC, SOT, MSS, and RMS-envelope similarity. All coefficients $\lambda_k$ are positive; distances enter $s_k$ with negative signs, and quality and similarity scores with positive signs, so a higher total reward is better. Reward weights and metric implementations appear in Appendices~\ref{app:rewards} and~\ref{app:metrics}, respectively.

For Stable Audio 3 data, $c$ denotes the associated text, and CLAP distance is the mean of audio--audio and text--audio cosine distances. In-domain data do not use $c$ and use only audio--audio distance. Text contributes to the reward, not to the program decoder input \citep{elizalde2023clap}.

The Basic Pitch reward compares hidden features preceding the note, onset, and contour prediction heads of a frozen model: cosine distances are computed frame by frame and averaged equally across time and the three branches, rather than comparing transcribed MIDI labels \citep{bittner2022basicpitch}. AudioBox CE and CU provide auxiliary feedback on sound enjoyment and usefulness, respectively \citep{tjandra2025audiobox}.

\paragraph{Group-relative policy optimization.}
For each target audio, we standardize the combined rewards within
its candidate group to obtain advantages $A_i$. We use a GRPO-style
group-relative policy-gradient update \citep{shao2024deepseekmath}:
\begin{equation}
\mathcal L_{\mathrm{GRPO}}
= -\frac{1}{G}\sum_{i=1}^{G} A_i \ell_i.
\label{eq:grpo}
\end{equation}
Here, $\ell_i$ is the sum of log probabilities at valid decision
positions under the sampling policy, and $A_i$ is treated as a
constant during optimization. Each fresh candidate group is used
once, without PPO clipping or reference-policy KL regularization.

Exploration uses $\epsilon$-greedy top-$k$ sampling, restricting random exploration to candidates considered plausible by the model. At each generation position, the $k$ highest-probability legal tokens are retained and their probabilities renormalized. Tokens below a probability threshold within this set are then removed. With probability $\epsilon$, a token is sampled according to the renormalized probabilities of the remaining candidates; with probability $1-\epsilon$, the highest-probability token is selected. This reduces unproductive exploration of low-probability values. Sampling and scoring use the same mixture distribution, including the greedy probability mass; hyperparameters and brief pseudocode appear in Appendix~\ref{app:hyperparameters}.

\section{Experiments}

We evaluate synthesizer inversion and text-driven generation separately. Both tasks focus on short sounds useful for music production rather than complete songs. Existing datasets rarely provide natural-language descriptions, audio, and complete synthesizer programs suitable for these tasks. We therefore use a synthetic audio test set oriented toward sound design and an FSD50K-based test set covering a broader range of real-world recordings, combining objective metrics with listening evaluations.

\subsection{Data and Experimental Setup}

\paragraph{Supervised data.}
We use Helm's 274 factory presets and one initialized state as seeds, expanding them into two million five-second stereo audio--program pairs through random parameter perturbations and varied MIDI performances. We manually design perturbation rules for different parameters, MIDI events, and routes to increase diversity while keeping the sound distribution close to practical creative settings. For example, performance probabilities depend on sound type: ambient presets are more likely to be played as chords, whereas bass presets are more likely to use single notes in a low register. The data contain single-note and chord performances rather than long melodies. Of the final samples, 90\% come from preset perturbations and 10\% from constrained random sampling.

\paragraph{Post-training data.}
We construct description templates and attribute combinations around sound-design constraints and programmatically generate 100,000 text descriptions. Stable Audio 3 generates audio with five random seeds per description, producing 500,000 clips. The descriptions are generated using GPT-6 and cover single notes, chords, short arpeggios, and sound textures while restricting full arrangements and long loops. During post-training, these clips are mixed with in-domain training audio at a $5:3$ ratio to improve both in-domain and out-of-domain performance. Generated audio serves only as a target sound and does not require target synthesizer-program annotations paired with the text.

\paragraph{Gemini--Woosh test set.}
To avoid dependence on the GPT text and Stable Audio 3 audio generators used during training, we construct a cross-generator test set using Gemini and Woosh. Gemini generates short sound-design descriptions from the same meta-prompt; deduplication yields 128 text prompts, which are passed to Woosh to generate reference sounds, each lasting five seconds. Synthesizer inversion uses these 128 Woosh waveforms, while text-driven generation uses the corresponding 128 descriptions.

\paragraph{FSD50K test set.}
We randomly select 256 recordings, each lasting five seconds, with usable sound descriptions from the official FSD50K evaluation split to evaluate the model across a broader range of real-world sounds \citep{fonseca2022fsd50k}.

\subsection{Synthesizer Inversion}

\paragraph{Baselines.}
We compare AutoSynth, Synth Permutations \citep{hayes2025permutations}, and DDSynth-RL \citep{wu2026ddsynth}. All three systems receive the same reference audio and render sounds using their own predicted MIDI and parameters. Inference settings appear in Appendix~\ref{app:baselines}.

\paragraph{Metrics.}
We report audio--audio CLAP cosine distance \citep{elizalde2023clap}, dynamic-time-warped MFCC distance (wMFCC), spectral optimal transport distance (SOT), multi-scale spectral distance (MSS), and RMS-envelope cosine similarity. These characterize matches in sound semantics, spectral envelopes, frequency-energy distributions, multi-scale time--frequency structure, and loudness envelopes, respectively. LanguageBind cosine distance between reference and generated audio provides a complementary measure of semantic alignment that is not used in reward optimization \citep{zhu2024languagebind}. Implementation details appear in Appendix~\ref{app:metrics}.

\begin{table}[t]
\caption{Synthesizer inversion on Gemini--Woosh and FSD50K. Bold denotes the best value for each metric within each test set.}
\label{tab:inversion}
\centering
\small
\renewcommand{\arraystretch}{0.88}
\setlength{\tabcolsep}{3pt}
\begin{tabular}{@{}lrrrrrr@{}}
\toprule
System & Audio CLAP $\downarrow$ & LanguageBind $\downarrow$ & wMFCC $\downarrow$ & SOT $\downarrow$ & RMS $\uparrow$ & MSS $\downarrow$ \\
\midrule
\multicolumn{7}{l}{\textit{Gemini--Woosh}} \\
\midrule
AutoSynth & \textbf{0.37} & \textbf{0.397} & \textbf{10.22} & \textbf{0.15} & 0.82 & \textbf{6.94} \\
Synth Permutations & 0.54 & 0.584 & 22.75 & 0.17 & 0.80 & 9.99 \\
DDSynth-RL & 0.49 & 0.482 & 10.34 & 0.16 & \textbf{0.87} & 7.04 \\
\midrule
\multicolumn{7}{l}{\textit{FSD50K}} \\
\midrule
AutoSynth & \textbf{0.67} & \textbf{0.649} & \textbf{14.95} & \textbf{0.16} & 0.632 & \textbf{13.00} \\
Synth Permutations & 0.69 & 0.688 & 23.34 & 0.18 & 0.627 & 15.26 \\
DDSynth-RL & 0.70 & 0.693 & 17.93 & 0.21 & \textbf{0.662} & 15.55 \\
\bottomrule
\end{tabular}
\end{table}

On Gemini--Woosh, AutoSynth has a substantial advantage in audio CLAP distance and slightly outperforms DDSynth-RL on wMFCC, SOT, and MSS, while achieving lower RMS-envelope similarity. Overall acoustic performance is relatively close. Broader sound-design coverage in the post-training data may help improve semantic alignment.

On FSD50K, the CLAP gap between AutoSynth and the other baselines narrows, possibly because the dataset contains many sounds outside the expressive range of the synthesizers, such as environmental recordings and speech. However, AutoSynth's lead on the acoustic metrics other than RMS becomes larger. This may reflect the greater expressive capacity afforded by modeling complete Helm programs, including MIDI, parameters, and routes. AutoSynth also leads on LanguageBind in both test sets, extending its alignment advantage beyond the CLAP reward.

\subsection{Text-Driven Generation}

\paragraph{Baselines and inference paths.}
We compare four synthesizer-program generation systems on the same texts: AutoSynth, Synth Permutations, DDSynth-RL, and CTAG \citep{cherep2024ctag}. The first three share the same intermediate Stable Audio 3 audio, while CTAG directly searches synthesizer controls from text. Stable Audio 3 and Woosh, which directly produce waveforms, are included only as end-to-end audio generation references.

\paragraph{Metrics.}
We report text--audio CLAP and LanguageBind distances, and AudioBox Content Enjoyment (CE), Content Usefulness (CU), and Production Quality (PQ). Neither LanguageBind nor PQ is optimized as a reward, enabling more objective baseline comparisons of semantic alignment and production quality, respectively. On FSD50K, we additionally compute FAD between generated audio and real recordings, and KL divergence between paired sound-event distributions \citep{tjandra2025audiobox,kilgour2018fad}. Metric definitions and generation settings appear in Appendices~\ref{app:metrics} and~\ref{app:baselines}, respectively.

\begin{table}[t]
\caption{Text-driven generation on Gemini and FSD50K descriptions, corresponding to the audio in Table~\ref{tab:inversion}.}
\label{tab:text-generation}
\centering
\small
\renewcommand{\arraystretch}{0.88}
\setlength{\tabcolsep}{3pt}
\begin{tabular}{@{}lrrrrrrr@{}}
\toprule
System & Text CLAP $\downarrow$ & LanguageBind $\downarrow$ & CE $\uparrow$ & CU $\uparrow$ & PQ $\uparrow$ & FAD $\downarrow$ & KL $\downarrow$ \\
\midrule
\multicolumn{8}{l}{\textit{Gemini}} \\
\midrule
AutoSynth & 0.550 & \textbf{0.638} & \textbf{3.759} & \textbf{7.23} & \textbf{7.433} & --- & --- \\
Synth Permutations & 0.642 & 0.718 & 3.409 & 6.88 & 7.372 & --- & --- \\
DDSynth-RL & 0.607 & 0.678 & 3.253 & 6.81 & 7.276 & --- & --- \\
CTAG & \textbf{0.519} & 0.645 & 3.020 & 6.58 & 6.888 & --- & --- \\
\midrule
\mbox{Stable Audio 3}$^*$ & 0.546 & 0.614 & 3.891 & 7.16 & 7.373 & --- & --- \\
Woosh$^*$ & 0.557 & 0.630 & 3.890 & 6.78 & 7.180 & --- & --- \\
\midrule
\multicolumn{8}{l}{\textit{FSD50K}} \\
\midrule
AutoSynth & 0.77 & 0.793 & \textbf{3.48} & \textbf{6.91} & 7.07 & \textbf{7.00} & 5.27 \\
Synth Permutations & 0.78 & 0.787 & 3.34 & 6.73 & \textbf{7.10} & 7.18 & 5.05 \\
DDSynth-RL & 0.80 & 0.809 & 2.87 & 6.57 & 6.89 & 10.26 & 5.29 \\
CTAG & \textbf{0.66} & \textbf{0.723} & 3.01 & 6.39 & 6.77 & 7.64 & \textbf{3.94} \\
\midrule
\mbox{Stable Audio 3}$^*$ & 0.50 & 0.573 & 3.62 & 6.60 & 6.93 & 2.73 & 1.84 \\
Woosh$^*$ & 0.48 & 0.591 & 3.58 & 6.15 & 6.78 & 2.05 & 1.69 \\
\bottomrule
\end{tabular}
\par\smallskip
\footnotesize $^*$End-to-end audio generation models, shown for reference rather than as synthesizer-program generation baselines.
\end{table}

On the sound-design-oriented Gemini descriptions, AutoSynth achieves lower text CLAP distance and higher CE, CU, and PQ than the other two inversion-based cascades. On both text-generation test sets, AutoSynth achieves higher PQ scores than Stable Audio 3 and Woosh. Since PQ is not directly optimized in the reward objective, this provides additional objective evidence of high production quality.

On Gemini, CLAP-guided CTAG has the lowest CLAP distance, but AutoSynth has a lower mean LanguageBind distance. Thus, CTAG's CLAP scores alone do not establish its alignment ability. AutoSynth is competitive on LanguageBind, which no evaluated method uses for training or search. CTAG's lower CE, CU, and PQ may reflect limitations of its synthesizer.

On FSD50K text-driven generation, AutoSynth does not achieve the best LanguageBind performance. This is consistent with the test set containing environmental sound descriptions beyond the current synthesizer's expressive range, indicating room to improve synthesizer-based sound matching for broader sound-effect generation tasks. Among the four synthesizer-program generation systems, AutoSynth still achieves the highest CU and CE.

\paragraph{Editability demonstration.}
Our demo shows GPT-6 editing generated synthesizer programs according to user instructions, followed by synthesizer re-rendering. The video is available on our \href{https://auto-synth.github.io/}{project page}.

\subsection{Subjective Evaluation}
\label{sec:listening}

Ten sound-creation professionals familiar with synthesizers rate both tasks on Gemini--Woosh and Freesound (FSD50K) for alignment with the reference audio or text and content quality. The protocol appears in Appendix~\ref{app:listening-protocol}.

\begin{figure}[!t]
\centering
\setlength{\abovecaptionskip}{3pt}
\includegraphics[width=.76\linewidth,trim=0 7bp 0 7bp,clip]{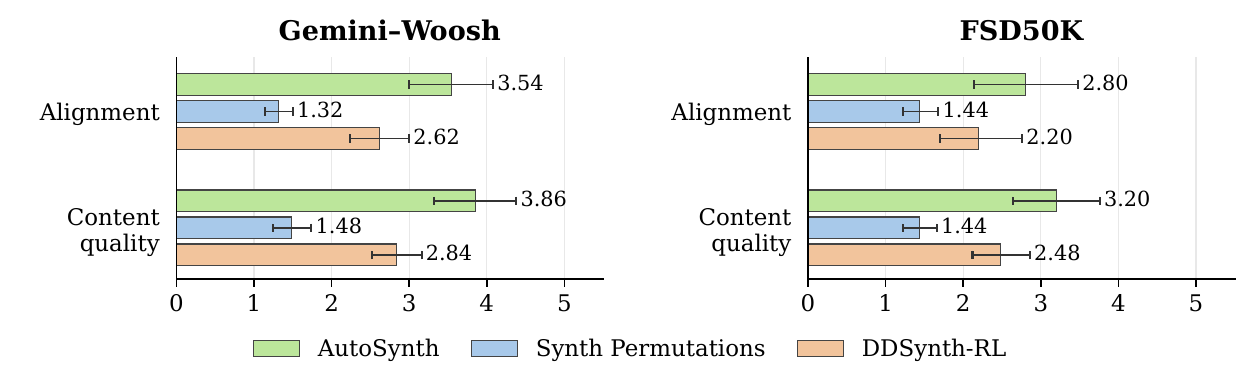}
\caption{Subjective synthesizer-inversion evaluation on Gemini--Woosh (left) and FSD50K (right). Error bars indicate 95\% bootstrap confidence intervals for the mean.}
\label{fig:subjective-inversion}
\includegraphics[width=.76\linewidth,trim=0 7bp 0 18bp,clip]{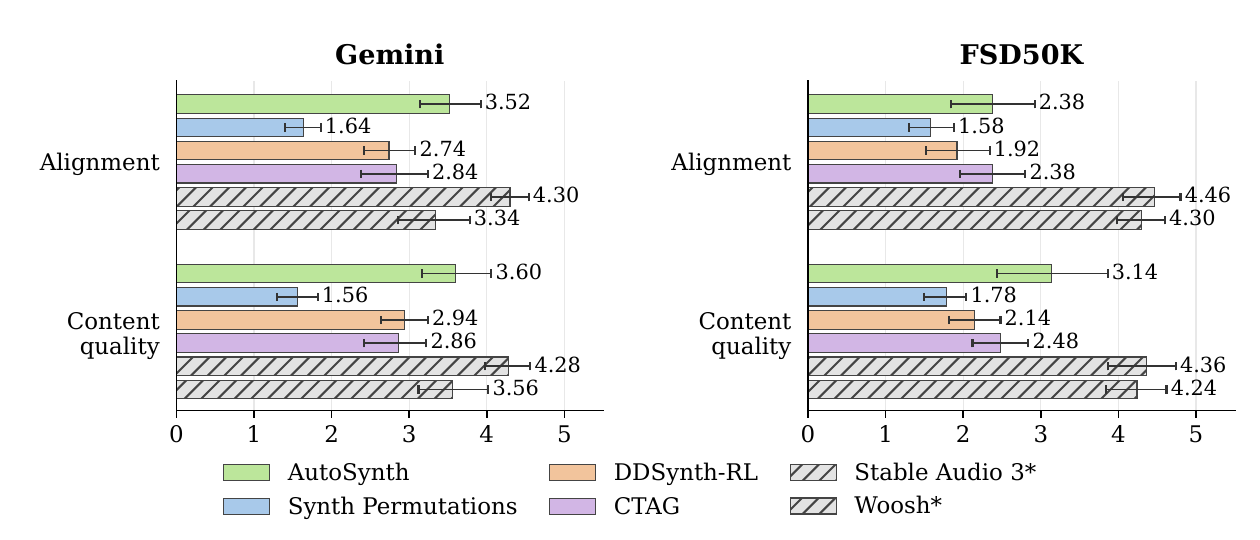}
\caption{Subjective text-driven generation evaluation on Gemini (left) and FSD50K (right). Gray, starred systems are end-to-end audio generation references, not synthesizer-program baselines. Error bars indicate 95\% bootstrap confidence intervals for the mean.}
\label{fig:subjective-text}
\end{figure}

\paragraph{Synthesizer inversion.}
AutoSynth leads both ratings on both test sets (Figure~\ref{fig:subjective-inversion}), supporting the objective results. Listeners prefer it despite DDSynth-RL's higher RMS similarity and similar Gemini--Woosh acoustic distances. Both methods receive lower FSD50K alignment ratings, consistent with recordings beyond their synthesizers' expressive range.

\paragraph{Text-driven generation.}
AutoSynth leads the synthesizer-program baselines on both dimensions for Gemini; on FSD50K, its mean content quality is highest and alignment ties CTAG (Figure~\ref{fig:subjective-text}). Despite CTAG's better CLAP scores, perceived alignment may depend on sound quality. AutoSynth's Gemini content-quality ratings are also comparable to those of end-to-end generators.

% Keep both listening figures together at the top of page 9.

\section{Discussion}
A unified benchmark is still needed for complete synthesizer-program generation. Differences in synthesizers' expressive capabilities and control structures make it difficult to disentangle the contributions of generation methods and underlying sound engines in cross-system comparisons. Methods that do not jointly model MIDI performances and variable-length modulation routes cannot directly support the joint Helm control space studied here with their existing program representations and output structures. AutoSynth's advantage lies not only in sound quality, but also in jointly modeling MIDI, timbral parameters, and modulation routes, providing a foundation for extending to more complex synthesizers.

In reinforcement learning for out-of-domain sound matching with black-box synthesizers, reward components and weights translate human sound-design preferences into training objectives. Sufficient supervised data from real creative settings can reflect these preferences through their distribution. However, program formats vary across synthesizers, and legally usable training data remain scarce, making data-augmentation rules and post-training reward design particularly important.

Sound design through synthesizer programs is a creative activity. In both inversion and generation, evaluation should consider creative intent and aesthetic preferences, which neither acoustic-feature fidelity nor semantic distance fully captures. Timbral aesthetics involve subtle perceptual differences that are difficult to verbalize and vary across listeners. Work such as AudioBox explores this issue, but a sound's appeal and usefulness depend on musical style and intended use, making context-free judgments difficult. Models therefore need more general prior knowledge of sound, music, and creative contexts to understand these aesthetics and generate synthesizer programs accordingly.

\section{Conclusion}
We present AutoSynth, which jointly models MIDI, synthesizer parameters, and variable-length modulation routes for synthesizer inversion and text-driven generation. Supervised learning and mixed-reward GRPO enable executable, editable sound controls without paired text--target-program annotations. Objective and subjective evaluations demonstrate a balance between input alignment and content quality in sound-design tasks, offering a practical path from audio generation to editable music creation.

\clearpage
\subsection*{Ethics Statement}
This research targets music production and sound design. We use the open-source Helm synthesizer and construct training data from its open-source presets and pretrained audio generation models. The use and redistribution of software, presets, models, and data follow their respective license requirements. Freesound source URLs, uploader names, and per-recording licenses have been retained in our data manifests. The listening study uses anonymous, voluntary participation with web-based informed consent before enrollment. All participants in the subjective listening evaluations provided informed consent.

\subsection*{Reproducibility Statement}
The model architecture, training and inference settings, program representation, and evaluation protocols are described in the main paper and appendices. We have open-sourced our training and evaluation code to support independent reproduction and comparison.

\subsection*{AI Use Statement}
Generative AI tools were used to assist with code implementation and checking, methodological discussion and interpretation of experimental results, and manuscript drafting and editing. They were also used to construct sound descriptions and organize metadata. Pretrained audio generation models supplied synthetic audio for training and evaluation, as described in the main paper.

\nocite{synthsetter}
\bibliographystyle{iclr2027_conference}
\bibliography{references}
\clearpage
\appendix
\makeatletter
\@addtoreset{table}{section}
\@addtoreset{figure}{section}
\makeatother
\renewcommand{\thetable}{\Alph{section}\arabic{table}}
\renewcommand{\thefigure}{\Alph{section}\arabic{figure}}
\numberwithin{equation}{section}
\renewcommand{\theequation}{\Alph{section}\arabic{equation}}
\section{Choice of Audio Encoder}
\label{app:encoders}

In addition to AutoSynth's jointly trained audio encoder, we consider freezing the encoder portions of Stable Audio 3's SAME-S and SAME-L \citep{parker2026same} and training synthesizer-program decoders on their representations. This comparison examines general-purpose audio compression representations and task-specific jointly learned representations, and explores the possibility of end-to-end text-driven generation by decoding directly from Stable Audio 3's latent space.

\begin{table}[h]
\caption{Encoder sizes. Counts exclude the program decoder and the SAME waveform decoder.}
\label{tab:encoders}
\centering\small
\begin{tabularx}{\linewidth}{lXrl}
\toprule
Audio encoder & Encoder architecture & Parameters & Frozen \\
\midrule
AutoSynth encoder & Stereo mel spectrograms, convolutional frontend, 12-layer audio Transformer & 28.02M & No \\
SAME-S & Waveform compression frontend, 6 backbone blocks of width 768 & 53.73M & Yes \\
SAME-L & Waveform compression frontend, 12 backbone blocks of width 1536 & 426.06M & Yes \\
\bottomrule
\end{tabularx}
\end{table}

The SAME conditioning branch projects the frozen encoder's 256-dimensional features to 512 dimensions with a trainable linear layer and supplies them to cross-attention in every program-decoder layer, without adding the audio Transformer encoder used in the main model.

\begin{figure}[ht]
\centering
\includegraphics[width=\linewidth]{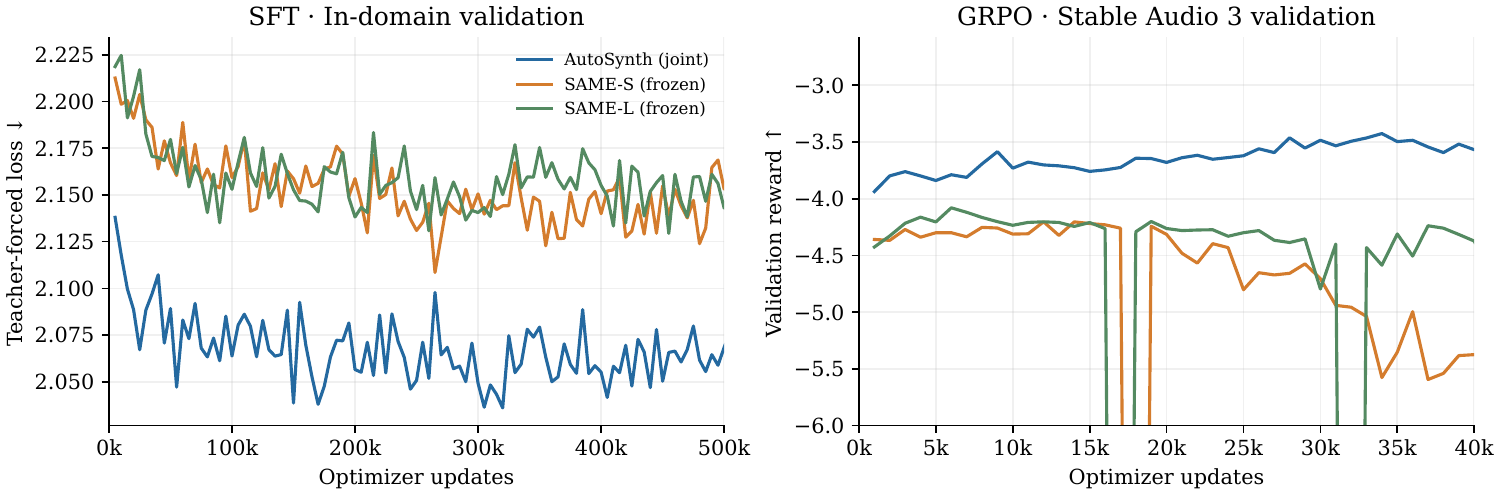}
\caption{Validation curves for three encoders across two training stages. Left: in-domain SFT loss, shown up to 500,000 updates. Right: Stable Audio 3 validation reward. Horizontal axes count optimizer updates within each stage.}
\label{fig:encoder-curves}
\end{figure}

\paragraph{Training-curve comparison.}
Figure~\ref{fig:encoder-curves} compares the three encoders during both training stages. During SFT, in-domain teacher-forced validation loss is evaluated every 5,000 updates. Post-training uses matching target mixtures, sampling strategies, and optimization settings, evaluating reward on the same 64 Stable Audio 3 validation clips every 1,000 updates.

The jointly trained encoder outperforms both frozen SAME branches on these validation measures. The frozen SAME branches also exhibit highly unstable post-training curves, with occasional collapses in validation reward.

\paragraph{Discussion.}
One possible explanation is that joint training adapts audio representations to predicting MIDI, timbre parameters, and modulation structure, whereas frozen general-purpose compression features need not organize information in a form convenient for these controls. Although the waveform-mediated cascade adds a decoding and encoding step, it can extract task-specific inversion features and is therefore not necessarily worse than directly reading generated latents. This is conceptually similar to using reference images to guide graphics generation.

\section{An Experiment in Creative Audio Super-Resolution}

Inversion can map reference audio containing noise or generation artifacts to sounds expressible by a native synthesizer and render them anew. This process can be studied as sound reconstruction or resynthesis, serving as a form of creative restoration or audio super-resolution in which the original sound is reinterpreted through a synthesizer program.

We use the same 128 Gemini--Woosh clips as in the main evaluation and independently apply three fixed corruptions: channel-independent Gaussian noise at 15~dB SNR, a sixth-order zero-phase Butterworth low-pass filter at 4~kHz, and symmetric hard clipping at a threshold 12~dB below the original peak. AutoSynth then inverts the corrupted audio to produce new renders.

Controls include the uncorrupted Woosh clips, corrupted clips, and AutoSynth resynthesis of uncorrupted inputs. Technical quality and content preservation are evaluated separately: AudioBox PQ estimates production quality, while CLAP distance to the uncorrupted reference assesses semantic-content preservation. CE and CU provide supplementary measures.

\begin{table}[ht]
\centering\small
\caption{Corrupted-input inversion and resynthesis, averaged over all 128 clips. CLAP distance uses the uncorrupted Woosh clip as reference; lower is better. Higher PQ is better.}
\label{tab:polisher}
\begin{tabular}{lrr}
\toprule
Condition & PQ $\uparrow$ & CLAP distance $\downarrow$ \\
\midrule
Uncorrupted Woosh & 7.18 & 0.000 \\
Uncorrupted input $\rightarrow$ AutoSynth & 7.43 & 0.372 \\
Noisy input & 5.07 & 0.306 \\
Noisy input $\rightarrow$ AutoSynth & 7.40 & 0.451 \\
Low-pass input & 7.10 & 0.134 \\
Low-pass input $\rightarrow$ AutoSynth & 7.34 & 0.375 \\
Clipped input & 6.69 & 0.158 \\
Clipped input $\rightarrow$ AutoSynth & 7.22 & 0.403 \\
\bottomrule
\end{tabular}
\end{table}

PQ increases by 2.33, 0.23, and 0.53 for noise, low-pass filtering, and clipping, respectively, while CLAP distance also increases in all three cases. The results are therefore better interpreted as creative resynthesis of corrupted sounds: projecting inputs into the synthesizer's sound space may produce editable sounds with higher estimated production quality, while changing their original timbral and performance details.

\section{SFT and GRPO Ablations}
\label{app:sft-grpo}

To assess post-training, we compare the SFT checkpoint used to initialize GRPO with the GRPO checkpoint used in the main evaluation. The comparison covers five task--dataset combinations: inversion of Gemini--Woosh, FSD50K, and in-domain Helm test audio, and generation from Gemini and FSD50K descriptions. Both stages use identical test examples and metric implementations; text-driven generation reuses the same intermediate Stable Audio 3 audio.

\begin{table}[ht]
\centering\small
\caption{SFT/GRPO comparison for synthesizer inversion.}
\label{tab:sft-grpo-inversion}
\setlength{\tabcolsep}{3pt}
\resizebox{\linewidth}{!}{%
\begin{tabular}{lrrrrrr}
\toprule
Test set & Audio CLAP $\downarrow$ & LanguageBind $\downarrow$ & wMFCC $\downarrow$ & SOT $\downarrow$ & RMS $\uparrow$ & MSS $\downarrow$\\
\midrule
Gemini--Woosh & 0.47 $\rightarrow$ 0.37 & 0.535 $\rightarrow$ 0.397 & 17.05 $\rightarrow$ 10.22 & 0.33 $\rightarrow$ 0.15 & 0.65 $\rightarrow$ 0.82 & 19.41 $\rightarrow$ 6.94\\
FSD50K & 0.73 $\rightarrow$ 0.67 & 0.746 $\rightarrow$ 0.649 & 22.05 $\rightarrow$ 14.95 & 0.31 $\rightarrow$ 0.16 & 0.472 $\rightarrow$ 0.632 & 24.33 $\rightarrow$ 13.00\\
Helm in-domain & 0.17 $\rightarrow$ 0.13 & 0.221 $\rightarrow$ 0.174 & 8.02 $\rightarrow$ 6.93 & 0.057 $\rightarrow$ 0.048 & 0.91 $\rightarrow$ 0.92 & 7.36 $\rightarrow$ 3.77\\
\bottomrule
\end{tabular}%
}
\end{table}

Compared with SFT alone, mixed-reward GRPO improves all six metrics on both out-of-domain test sets and the in-domain Helm test set, including LanguageBind, which is not used in the reward. Thus, in this evaluation, improved out-of-domain inversion does not come at the expense of in-domain reconstruction.

\begin{table}[H]
\centering\small
\caption{SFT/GRPO comparison for text-driven generation.}
\label{tab:sft-grpo-text}
\setlength{\tabcolsep}{3pt}
\resizebox{\linewidth}{!}{%
\begin{tabular}{lrrrrrrr}
\toprule
Test set & Text CLAP $\downarrow$ & LanguageBind $\downarrow$ & CE $\uparrow$ & CU $\uparrow$ & PQ $\uparrow$ & FAD $\downarrow$ & KL $\downarrow$\\
\midrule
Gemini & 0.601 $\rightarrow$ 0.550 & 0.671 $\rightarrow$ 0.638 & 3.469 $\rightarrow$ 3.759 & 6.92 $\rightarrow$ 7.23 & 7.278 $\rightarrow$ 7.433 & --- & ---\\
FSD50K & 0.768 $\rightarrow$ 0.771 & 0.776 $\rightarrow$ 0.793 & 3.24 $\rightarrow$ 3.48 & 6.73 $\rightarrow$ 6.91 & 6.97 $\rightarrow$ 7.07 & 8.83 $\rightarrow$ 7.00 & 5.04 $\rightarrow$ 5.27\\
\bottomrule
\end{tabular}%
}
\end{table}

\paragraph{Text-driven generation results.}
With the same Gemini descriptions and intermediate Stable Audio 3 audio, GRPO reduces both text CLAP and LanguageBind distances while improving CE, CU, and PQ. On FSD50K, all three AudioBox scores and FAD also improve, although text CLAP distance, LanguageBind distance, and KL become worse. This indicates that, on the more general test set, post-training primarily improves the quality and usefulness scores of text-driven outputs.

Overall, mixed-reward GRPO improves most metrics across the test sets for both tasks, supporting the contribution of rendering-based post-training to model performance.

\section{Program Representation and Autoregressive Ordering}
\label{app:ordering}

The current sequence is organized as follows:
\begin{quote}
\ttfamily\small
BOS\\
$\rightarrow$ [NOTE, pitch, velocity, onset, duration] $\times N$\\
$\rightarrow$ NOTE\_END\\
$\rightarrow$ [controller\_type, value, time] $\times C$\\
$\rightarrow$ CONTROL\_END\\
$\rightarrow$ fixed\_parameter\_1 $\rightarrow\cdots\rightarrow$ fixed\_parameter\_133\\
$\rightarrow$ [source, destination, amount] $\times K$\\
$\rightarrow$ ROUTE\_END
\end{quote}

Here, $1\leq N\leq4$, $0\leq C\leq4$, and $0\leq K\leq13$. Controller events cover modulation-wheel and aftertouch information supported by the implementation. Fixed parameters include oscillator, filter, envelope, effect, and performance settings; their ordering is determined by a versioned parameter schema. The native schema contains 139 fixed parameters; the current task fixes six delay-related parameters to a disabled configuration and excludes them from prediction.

Modulation routes are source--destination--amount triplets, canonicalized by a fixed source and destination ordering. Continuous parameters and modulation amounts use 128 quantization levels; discrete parameters retain their native categories. End markers allow different numbers of events and routes across samples.

The current MIDI $\rightarrow$ fixed parameters $\rightarrow$ modulation routes order first specifies the performance, then configures the timbre, and finally predicts modulation relationships. Within fixed parameters, structural choices and switches precede their dependent values. For example, oscillator waveform precedes tuning and volume, and effect switches precede their control parameters. Final output volume is placed last.

\paragraph{Ordering ablation.}
We compare three sequences: (1) the current order; (2) a fixed random permutation of the 133 parameters, leaving MIDI and routes in place; and (3) fixed parameters $\rightarrow$ MIDI $\rightarrow$ modulation routes, retaining the canonical internal parameter order. The second condition uses one shared permutation for all samples, rather than reshuffling each sample. Both new SFT runs are trained for 100,000 steps and compared with the current order using in-domain validation loss at matching steps. Data, model size, optimization settings, and validation samples are held constant.

\begin{figure}[H]
\centering
\includegraphics[width=\linewidth]{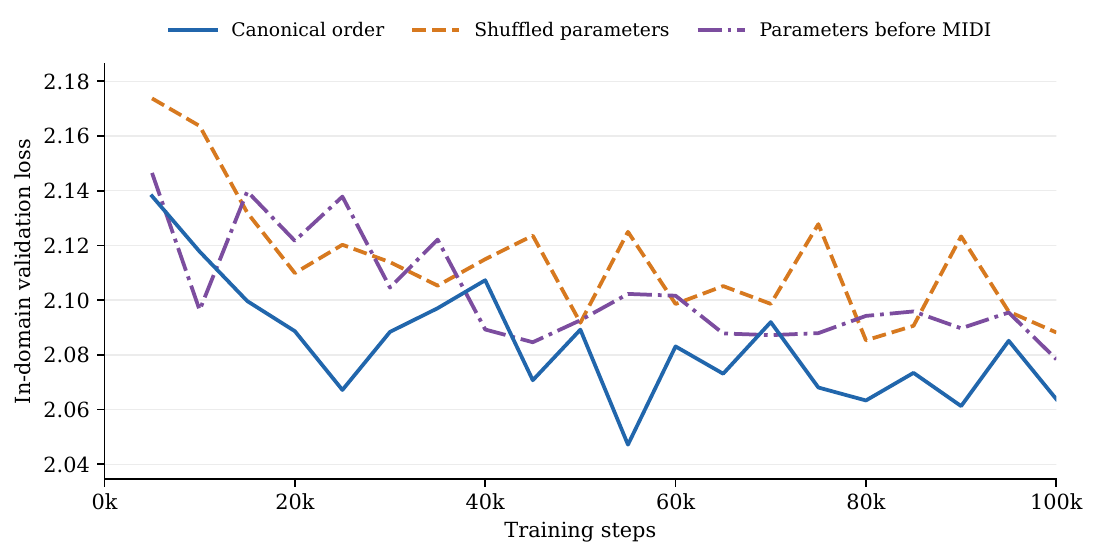}
\caption{In-domain validation loss for different autoregressive orders over the first 100,000 steps.}
\label{fig:order-val-loss}
\end{figure}

Figure~\ref{fig:order-val-loss} shows that the current order generally achieves lower in-domain validation loss, indicating that autoregressive ordering affects training performance. The final order was determined through sound-design experience and preliminary experiments; the complete parameter order is listed below.

\begingroup\small
\begin{longtable}{@{}>{\raggedright\arraybackslash}p{.09\linewidth}>{\raggedright\arraybackslash}p{.20\linewidth}>{\raggedright\arraybackslash}p{.65\linewidth}@{}}
\caption{Autoregressive order of fixed parameters, read left to right within each row.}\label{tab:parameter-order}\\
\toprule
Position & Group & Parameters in generation order\\
\midrule
\endfirsthead
\multicolumn{3}{l}{\tablename\ \thetable\ (continued)}\\
\toprule
Position & Group & Parameters in generation order\\
\midrule
\endhead
\bottomrule
\endfoot
1--7 & Global performance & \path{beats_per_minute}, \path{polyphony}, \path{legato}, \path{portamento_type}, \path{portamento}, \path{pitch_bend_range}, \path{velocity_track} \\
8--14 & Oscillator 1 & \path{osc_1_waveform}, \path{osc_1_transpose}, \path{osc_1_tune}, \path{osc_1_unison_voices}, \path{osc_1_unison_detune}, \path{unison_1_harmonize}, \path{osc_1_volume} \\
15--21 & Oscillator 2 & \path{osc_2_waveform}, \path{osc_2_transpose}, \path{osc_2_tune}, \path{osc_2_unison_voices}, \path{osc_2_unison_detune}, \path{unison_2_harmonize}, \path{osc_2_volume} \\
22--30 & Cross-modulation and sources & \path{cross_modulation}, \path{sub_waveform}, \path{sub_octave}, \path{sub_shuffle}, \path{sub_volume}, \path{noise_volume}, \path{osc_feedback_amount}, \path{osc_feedback_transpose}, \path{osc_feedback_tune} \\
31--34 & Amplitude envelope & \path{amp_attack}, \path{amp_decay}, \path{amp_sustain}, \path{amp_release} \\
35--47 & Filter and envelope & \path{filter_on}, \path{filter_style}, \path{filter_shelf}, \path{filter_blend}, \path{filter_drive}, \path{cutoff}, \path{resonance}, \path{keytrack}, \path{fil_env_depth}, \path{fil_attack}, \path{fil_decay}, \path{fil_sustain}, \path{fil_release} \\
48--51 & Modulation envelope & \path{mod_attack}, \path{mod_decay}, \path{mod_sustain}, \path{mod_release} \\
52--57 & Monophonic LFO 1 & \path{mono_lfo_1_waveform}, \path{mono_lfo_1_retrigger}, \path{mono_lfo_1_sync}, \path{mono_lfo_1_frequency}, \path{mono_lfo_1_tempo}, \path{mono_lfo_1_amplitude} \\
58--63 & Monophonic LFO 2 & \path{mono_lfo_2_waveform}, \path{mono_lfo_2_retrigger}, \path{mono_lfo_2_sync}, \path{mono_lfo_2_frequency}, \path{mono_lfo_2_tempo}, \path{mono_lfo_2_amplitude} \\
64--68 & Polyphonic LFO & \path{poly_lfo_waveform}, \path{poly_lfo_sync}, \path{poly_lfo_frequency}, \path{poly_lfo_tempo}, \path{poly_lfo_amplitude} \\
69--74 & Step-sequencer settings & \path{num_steps}, \path{step_sequencer_retrigger}, \path{step_sequencer_sync}, \path{step_frequency}, \path{step_sequencer_tempo}, \path{step_smoothing} \\
75--90 & Step values 0--15 & \path{step_seq_00}, \path{step_seq_01}, \path{step_seq_02}, \path{step_seq_03}, \path{step_seq_04}, \path{step_seq_05}, \path{step_seq_06}, \path{step_seq_07}, \path{step_seq_08}, \path{step_seq_09}, \path{step_seq_10}, \path{step_seq_11}, \path{step_seq_12}, \path{step_seq_13}, \path{step_seq_14}, \path{step_seq_15} \\
91--106 & Step values 16--31 & \path{step_seq_16}, \path{step_seq_17}, \path{step_seq_18}, \path{step_seq_19}, \path{step_seq_20}, \path{step_seq_21}, \path{step_seq_22}, \path{step_seq_23}, \path{step_seq_24}, \path{step_seq_25}, \path{step_seq_26}, \path{step_seq_27}, \path{step_seq_28}, \path{step_seq_29}, \path{step_seq_30}, \path{step_seq_31} \\
107--113 & Arpeggiator & \path{arp_on}, \path{arp_pattern}, \path{arp_octaves}, \path{arp_sync}, \path{arp_frequency}, \path{arp_tempo}, \path{arp_gate} \\
114--117 & Distortion & \path{distortion_on}, \path{distortion_type}, \path{distortion_drive}, \path{distortion_mix} \\
118--120 & Formant & \path{formant_on}, \path{formant_x}, \path{formant_y} \\
121--124 & Reverb & \path{reverb_on}, \path{reverb_damping}, \path{reverb_feedback}, \path{reverb_dry_wet} \\
125--132 & Stutter & \path{stutter_on}, \path{stutter_sync}, \path{stutter_frequency}, \path{stutter_tempo}, \path{stutter_softness}, \path{stutter_resample_sync}, \path{stutter_resample_frequency}, \path{stutter_resample_tempo} \\
133 & Output volume & \path{volume} \\
\end{longtable}
\endgroup

\raggedbottom
\section{Implementation and Evaluation Details}

\label{app:implementation}
\subsection{Model and Training Hyperparameters}
\label{app:hyperparameters}

\begin{table}[H]
\caption{Main model and optimization settings.}
\label{tab:hyperparameters}
\centering\small
\begin{tabularx}{\linewidth}{lX}
\toprule
Item & Setting \\
\midrule
Audio input & 44.1 kHz, 5 seconds, stereo \\
Mel spectrograms & 128 bands per channel, 25 ms Hamming window, 100 frames/s \\
Encoder / decoder layers & 12 / 12 \\
Hidden width / heads / FFN width & 512 / 8 / 1024 \\
Normalization and activation & Pre-LayerNorm, GELU, dropout 0.1 \\
Value output layer & Shared Linear(512, 128), with field-specific legal categories \\
SFT optimizer & AdamW \citep{loshchilov2019adamw}; learning rate $2\times10^{-4}$; weight decay $10^{-4}$ \\
SFT learning-rate schedule & 10,000-step warmup followed by cosine decay \\
SFT batch size & 16 \\
SFT classification-logit temperature & 0.2 \\
Ordered-field Gaussian kernel $\sigma$ & 0.02 \\
GRPO optimizer & AdamW; learning rate $5\times10^{-6}$; weight decay 0 \\
Targets per update / candidates per target & 3 / 8 \\
Updates per candidate group & 1 \\
Sampling temperature / $\epsilon$ / top-$k$ & 1.0 / 0.7 / 16 \\
Probability threshold after top-$k$ normalization & 0.1 \\
Gradient-norm clipping & 1.0 \\
GRPO validation interval & 1,000 steps \\
\bottomrule
\end{tabularx}
\end{table}

The probability threshold is applied after renormalizing within the top-$k$ set. If no token passes the threshold, the implementation selects the highest-probability token.

\paragraph{Sampling and scoring.}
The pseudocode below applies at each generation position. Here, $h$ contains the audio condition and generated prefix, $T$ is the temperature, and $\tau$ is the probability threshold. \texttt{top\_k} sets unretained probabilities to zero. Scoring reconstructs $q$ at the same prefix and reads the sampled token's probability without resampling.
\begin{quote}\small
\begin{tabular}{@{}l@{}}
$p\leftarrow\operatorname{softmax}(\operatorname{mask\_legal}(\operatorname{logits}(h))/T)$\\
$g\leftarrow\operatorname{argmax}(p)$\\
$p\leftarrow\operatorname{normalize}(\operatorname{top\_k}(p,k))$\\
$p[p<\tau]\leftarrow0$\\
\textbf{if} $\sum_a p[a]>0$: $q\leftarrow\epsilon\operatorname{normalize}(p)+(1-\epsilon)\operatorname{one\_hot}(g)$\\
\textbf{else}: $q\leftarrow\operatorname{one\_hot}(g)$\\
$a\sim\operatorname{Categorical}(q)$\\
$\mathrm{log\_prob}\leftarrow\log(\max(q[a],10^{-12}))$
\end{tabular}
\end{quote}

\subsection{Mixed-Reward Weights}

\label{app:rewards}

\begin{table}[H]
\caption{Coefficients used in Equation~\eqref{eq:reward}. All coefficients are positive; each metric's direction determines its sign in the reward.}
\label{tab:weights}
\centering\small
\begin{tabular}{llr}
\toprule
Component & Direction & Coefficient \\
\midrule
CLAP distance & Negative & 8 \\
Basic Pitch hidden-feature distance & Negative & 12 \\
AudioBox CE, normalized to $[0,1]$ & Positive & 1 \\
AudioBox CU, normalized to $[0,1]$ & Positive & 1 \\
wMFCC & Negative & 0.07 \\
SOT & Negative & 3.58 \\
MSS & Negative & 0.11 \\
RMS-envelope similarity & Positive & 0.24 \\
\bottomrule
\end{tabular}
\end{table}

The three Basic Pitch branches each receive weight $1/3$ before their mean is multiplied by the overall coefficient 12.

\subsection{Evaluation Metrics}
\label{app:metrics}

\begin{itemize}
\item \textbf{CLAP:} We use the paired MS-CLAP 2023 audio and text encoders and compute $1-\cos(\cdot,\cdot)$ without the contrastive-logit temperature factor. Evaluation uses FP32 with TF32 disabled.
\item \textbf{LanguageBind:} We use the frozen official LanguageBind Audio\_FT model and its paired text encoder \citep{zhu2024languagebind} to extract 768-dimensional projected features, computing $1-\cos(\cdot,\cdot)$ without logit scaling. Inversion uses audio--audio distance, whereas text-driven generation uses text--audio distance. Audio is averaged across channels and processed using the official 16 kHz resampling, filterbank, and normalization pipeline; text-generation evaluation retains the existing five-second audio preprocessing. All methods use identical test examples and original English descriptions. LanguageBind is not used in reward optimization.
\item \textbf{wMFCC:} We use 20-dimensional MFCCs with time retained as the frame axis. The local cost matrix uses the mean L1 distance over coefficients, and the result is the normalized dynamic-time-warping distance with the \texttt{symmetric2} step pattern.
\item \textbf{SOT:} STFT magnitudes from 50 ms windows and 20 ms hops are normalized framewise. We compute the one-dimensional Wasserstein distance between frequency distributions and average over frames.
\item \textbf{MSS:} We compute mean absolute differences between dB mel spectrograms at three scales, using window lengths of 10/25/100 ms, hops of 5/10/50 ms, and 32/64/128 mel bands, respectively, then average over scales.
\item \textbf{RMS:} RMS envelopes are extracted using 50 ms windows and 25 ms hops and compared by cosine similarity.
\item \textbf{AudioBox:} Evaluation tables report raw predicted CE, CU, and PQ scores; only reward CE and CU are normalized to $[0,1]$. CE and CU contribute to reward optimization, while PQ is used only to evaluate production quality.
\item \textbf{FAD:} We use 128-dimensional VGGish features \citep{hershey2017cnn} without PCA or quantization. Features from 0.96-second windows across clips are pooled, and a set-level Fr\'echet distance is computed from the means and covariances of all reference and generated features, rather than averaging purported per-clip FAD values.
\item \textbf{KL:} A frozen PaSST sound-event classifier \citep{koutini2022passt} produces 527-dimensional logits, followed by softmax. We compute $D_{\mathrm{KL}}(p_{\mathrm{reference}}\Vert p_{\mathrm{generated}})$ for each pair and average across pairs using natural logarithms. The implementation follows \href{https://github.com/kkoutini/PaSST}{PaSST} and the \href{https://github.com/haoheliu/audioldm_eval/tree/passt_replace_panns}{AudioLDM evaluation toolkit}.
\end{itemize}

\subsection{Baseline Methods and Inference Settings}
\label{app:baselines}

\paragraph{Stable Audio 3.}
This method performs conditional diffusion in the latent space of the SAME audio autoencoder, supporting text-driven music and sound-effect generation with controllable output duration \citep{evans2026stable}. We use Medium weights, eight generation steps, and CFG 1 to provide identical intermediate audio for the text-generation paths of AutoSynth, Synth Permutations, and DDSynth-RL. Its directly generated waveforms also serve as an end-to-end audio generation reference, rather than a synthesizer-program generation baseline.

\paragraph{Woosh.}
Woosh targets sound-effect generation. Its Woosh-Flow model uses flow matching to generate audio latents conditioned on text, which an audio decoder converts into waveforms \citep{hadjeres2026woosh}. We use the official open-source Woosh-Flow-Public release both to construct reference audio for the Gemini--Woosh inversion test set and to generate audio from descriptions in the two text test sets as an end-to-end generation reference. Its outputs are waveforms rather than synthesizer programs.

\paragraph{Synth Permutations.}
This method models inversion as audio-conditioned parameter generation and uses approximately permutation-equivariant flow matching to handle parameter-space symmetries, representing multiple valid parameter configurations for the same sound \citep{hayes2025permutations}. Since the original work did not provide pretrained checkpoints, we retrain a Surge XT inversion model using its open-source implementation. Inference uses 100 RK4 integration steps and CFG 2. The model receives reference audio for inversion and Stable Audio 3-generated intermediate audio for text-driven generation, with predictions rendered through Surge XT.

\paragraph{DDSynth-RL.}
This method uses masked discrete diffusion to iteratively recover discretized synthesizer controls conditioned on reference audio. GRPO-style post-training with rendered-audio rewards improves sound matching without backpropagating through the native synthesizer \citep{wu2026ddsynth}. We use the Dexed checkpoint with the highest validation reward after fine-tuning with a joint CREPE and CLAP reward, retaining greedy value selection and normalized-entropy-based position selection. The model inverts test-set reference audio or Stable Audio 3 intermediate audio, and Dexed renders the outputs for comparison in the two tasks.

\paragraph{CTAG.}
This method uses text--audio similarity to guide evolutionary search over a modular synthesizer with 78 parameters, producing parameter configurations that can be inspected and edited \citep{cherep2024ctag}. We use it for text-driven generation, searching directly from each test description without intermediate Stable Audio 3 audio. Following the \href{https://github.com/PapayaResearch/ctag}{official search configuration}, we use the SynthAX Voice synthesizer, Learned Evolution Strategy (LES), a population of 50, and 300 iterations. Search uses LAION-CLAP, with one independent run per description and the best search candidate returned as output.

\section{Listening Evaluation Protocol}
\label{app:listening-protocol}

We invite ten participants with backgrounds in sound creation and familiarity with synthesizers to an anonymous, blinded listening study. The questionnaire contains four sections covering two tasks and two test sets. Each section has five randomly sampled questions, giving twenty questions per participant. Synthesizer inversion presents reference audio and candidates from three systems; text-driven generation presents a description and candidates from six systems. Method names are hidden and candidate order is independently randomized for each question.

Participants separately rate alignment and content quality using the questions below, on integer scales from 1 to 5, with higher scores indicating better performance.

\begin{table}[H]
\centering\small
\begin{tabularx}{\linewidth}{@{}>{\raggedright\arraybackslash}p{.17\linewidth}>{\raggedright\arraybackslash}p{.29\linewidth}>{\raggedright\arraybackslash}X@{}}
\toprule
Dimension & Question & Evaluation criteria\\
\midrule
Alignment: synthesizer inversion & Does this sound faithfully reproduce the reference audio? & Consider timbre, pitch, articulation, and changes over time; 1 means not at all and 5 means very faithfully.\\
Alignment: text-driven generation & Does this sound faithfully follow the text description? & Consider the described timbre, pitch, articulation, and temporal changes; 1 means not at all and 5 means very faithfully.\\
Content quality: both tasks & How good is this sound as material for music production or sound design? & Consider audio quality, unintended artifacts, and creative usefulness; 1 means very poor and 5 means excellent. Intentional distortion is not a defect, and quality is rated independently of alignment.\\
\bottomrule
\end{tabularx}
\end{table}

Participants are asked to use headphones in a quiet environment. Audio is not loudness-normalized, and participants may adjust playback volume to avoid bias from loudness differences. The current results contain 1,800 valid scores, with 50 ratings per method, test set, and dimension.

\end{document}